\documentclass[aps,prl,superscriptaddress,showpacs,twocolumn]{revtex4-1}

\usepackage{amssymb,amsfonts,amsmath}
\usepackage{graphicx}
\usepackage{dcolumn}
\usepackage{bm}
\usepackage{comment}
\usepackage[dvipsnames]{xcolor}
\usepackage{hyperref}

\usepackage[
margin=0.7in,
]{geometry}

\usepackage{bm}
\usepackage{physics}
\usepackage{graphicx}
\hypersetup{
	colorlinks=true,       
	linkcolor=blue,        
	citecolor=blue,        
	filecolor=magenta,     
	urlcolor=blue         
}

\usepackage{comment}
\usepackage[dvipsnames]{xcolor} 
\usepackage{blindtext}

\newcommand{\vc}[1]{\bm{\mathrm{#1}}}
\newcommand{\mt}[1]{\mathrm{#1}}
\newcommand{\bgpl}{\big ( \,}
\newcommand{\bgpr}{\, \big )}

\newcommand{\bgbl}{\big [ \,}
\newcommand{\bgbr}{\, \big ]}

\newcommand{\Bgpl}{\Big ( \,}
\newcommand{\Bgpr}{\, \Big )}
\newcommand{\Bggpl}{\Bigg ( \,}
\newcommand{\Bggpr}{\, \Bigg )}
\newcommand{\Bgbl}{\Big [ \,}
\newcommand{\Bgbr}{\, \Big ]}
\newcommand{\Bggbl}{\Bigg [ \,}
\newcommand{\Bggbr}{\, \Bigg ]}

\definecolor{Nathanorange}{rgb}{1,0.498,0.0549}

\begin{document}

\preprint{APS/123-QED}

\title{Quantized transport of solitons in nonlinear Thouless pumps: \\ From Wannier drags to ultracold topological mixtures}

\author{N. Mostaan}
\email{Nader.Mostaan@physik.uni-muenchen.de}
 \affiliation{Department of Physics and Arnold Sommerfeld Center for Theoretical Physics (ASC), Ludwig-Maximilians-Universität München, Theresienstr. 37, D-80333 München, Germany}
 \affiliation{Munich Center for Quantum Science and Technology (MCQST), Schellingstr. 4, D-80799 München, Germany}
 \affiliation{CENOLI, Universit\'e Libre de Bruxelles, CP 231, Campus Plaine, B-1050 Brussels, Belgium}
 \author{F. Grusdt}
 \affiliation{Department of Physics and Arnold Sommerfeld Center for Theoretical Physics (ASC), Ludwig-Maximilians-Universität München, Theresienstr. 37, D-80333 München, Germany}
 \affiliation{Munich Center for Quantum Science and Technology (MCQST), Schellingstr. 4, D-80799 München, Germany}

\author{N. Goldman}
\email{ngoldman@ulb.ac.be}
\affiliation{CENOLI, Universit\'e Libre de Bruxelles, CP 231, Campus Plaine, B-1050 Brussels, Belgium}

\date{\today}

\begin{abstract}
Recent progress in synthetic lattice systems has opened the door to novel explorations of topological matter. In particular, photonic devices and ultracold matter waves offer the unique possibility of studying the rich interplay between topological band structures and tunable nonlinearities. In this emerging field of nonlinear topological physics, a recent experiment revealed the quantized motion of localized nonlinear excitations (solitons) upon driving a Thouless pump sequence; the reported observations suggest that the quantized displacement of solitons is dictated by the Chern number of the band from which they emanate. In this work, we elucidate the origin of this intriguing nonlinear topological effect, by showing that the motion of solitons is established by the quantized displacement of Wannier functions. Our general theoretical approach, which fully clarifies the central role of the Chern number in solitonic pumps, provides a rigorous framework for describing the topological transport of nonlinear excitations in a broad class of physical systems. Exploiting this interdisciplinarity, we introduce an interaction-induced topological pump for ultracold atomic mixtures, where solitons of impurity atoms experience a quantized drift resulting from genuine interaction processes with their environment. 
\end{abstract}

\maketitle

\subsection{\label{sec:level0}Introduction}

Quantized responses have been a central theme throughout the realm of topological physics, which was initiated with the discovery of the quantum Hall effects in two-dimensional electron gases~\cite{hasan2010colloquium,qi2011topological}. A wide variety of topological band structures have been revealed over the last decades, leading to the identification of various forms of quantized responses, from quantized Faraday and Kerr rotations in three-dimensional topological insulators~\cite{wu2016quantized} to quantized circular dichroism~\cite{asteria2019measuring} and topological Bloch oscillations~\cite{li2016bloch,hoeller2018topological} in two-dimensional ultracold atomic gases. An emblematic and minimal instance of quantized topological transport concerns the adiabatic motion of a quantum particle moving in a slowly-varying periodic potential, an effect known as the Thouless pump~\cite{PhysRevB.27.6083}. In this setting, the center-of-mass motion is quantized according to the Chern number of the underlying band structure, as defined over a hybrid momentum-time space~\cite{xiao2010berry}. The realization of synthetic lattice systems has allowed for the experimental implementation of Thouless pumps and for the observation of the related quantized motion, in both photonics~\cite{kraus2012topological,verbin2015topological,zilberberg2018photonic,grinberg2020robust,cerjan2020thouless,ke2016topological} and ultracold gases setups~\cite{lohse2016thouless,nakajima2016topological}.

Interestingly, synthetic topological systems~\cite{ozawa2019topological,cooper2019topological} can operate beyond the linear regime of the Schrödinger equation, hence opening the door to nonlinear topological physics~\cite{smirnova2020nonlinear}. In this emerging framework, a central topic concerns the possible interplay between nonlinear excitations, known as solitons, and the underlying topological band structure~\cite{leykam2016edge,solnyshkov2017chirality,st2017lasing,bisianov2019stability,ivanov2020vector,gonzalez2020dynamical,mukherjee2020observation,mukherjee2020observation2,xia2020nontrivial,pernet2021topological,mittal2021topological,kirsch2021nonlinear}. Interestingly, exact correspondences between topological indices and nonlinear modes have been identified in mechanical systems~\cite{lo2021topology} and for the Korteweg-de-Vries equation of fluid dynamics~\cite{oblak2020berry}, hence allowing for a formal topological classification of nonlinear excitations in certain special cases. In the context of nonlinear topological photonics, a recent experimental study reported on the quantized motion of solitons in a lattice system undergoing a Thouless pump sequence~\cite{jurgensen2021quantized}. 
Despite the presence of considerable nonlinearity, these observations suggest that the quantization of the solitons displacement is dictated by the Chern numbers of the underlying band structure.

\begin{figure}[h!]
\centering
\includegraphics[width=6.7cm]{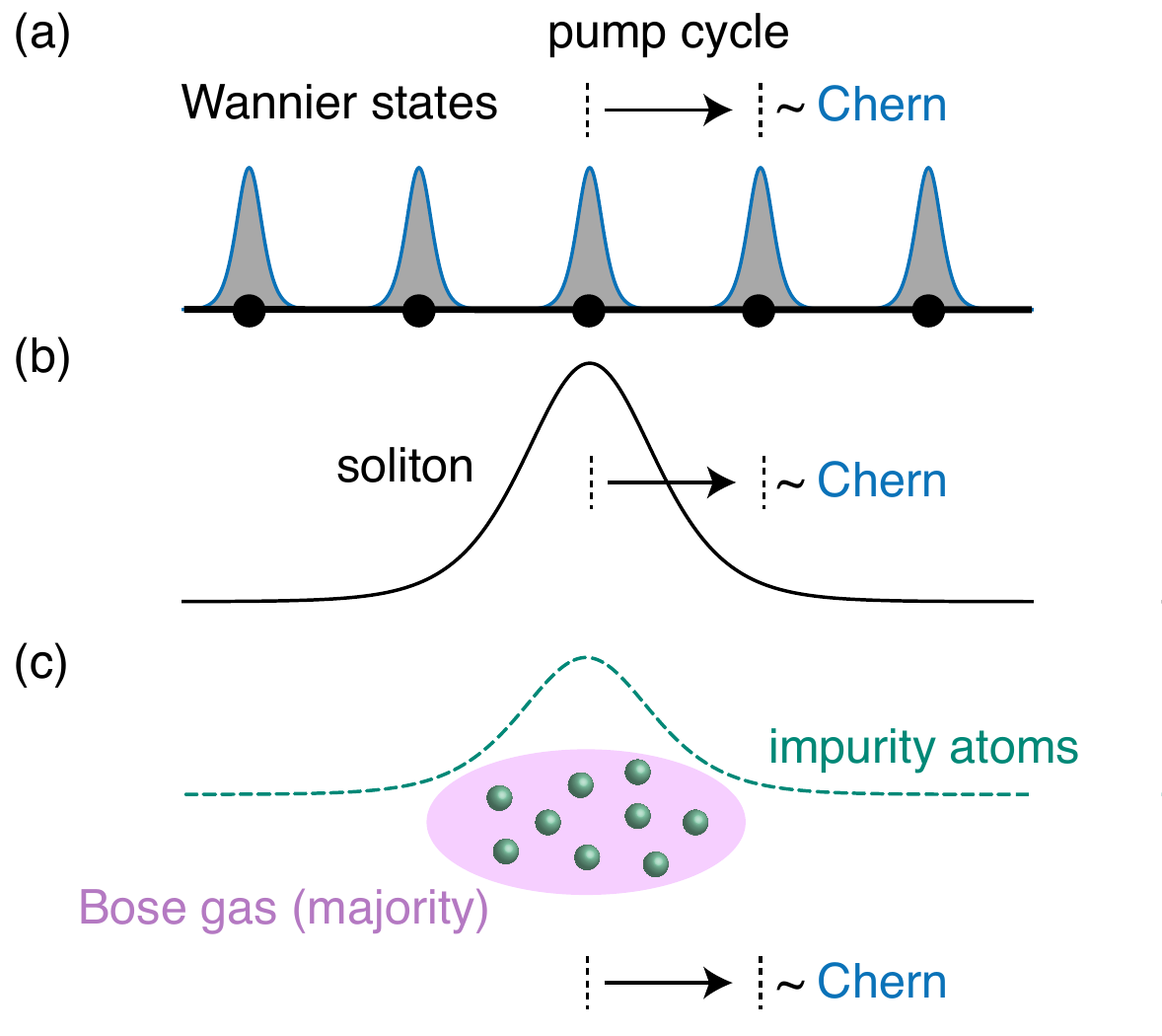}
\vspace{-0.25cm}
\caption{(a) In a Thouless pump, the Wannier functions perform a quantized drift established by the Chern number of the corresponding band. (b) In a nonlinear setting, the motion of a soliton follows the quantized Wannier drift. (c) In a Bose-Bose atomic mixture, quantized pumping can be induced by interactions :~a soliton of impurity atoms is dragged by the driven majority atoms, leading to interaction-induced topological transport.}
\label{Fig_sketch}
\end{figure}

It is the aim of this work to elucidate and explore the quantized transport of solitons in nonlinear topological Thouless pumps. Inspired by the experiment of Ref.~\cite{jurgensen2021quantized}, we address this topic by considering a general class of one-dimensional systems described by the discrete nonlinear Schr\"odinger equation (DNLS)~\cite{Sulem, kevrekidis2009discrete}. In the regime of weak nonlinearity, solitons are known to predominantly occupy a single Bloch band~\cite{kevrekidis2009discrete}; following Ref.~\cite{alfimov2002wannier}, we represent the solitons in terms of maximally localized Wannier functions, defined in the corresponding Bloch band. In this Wannier representation, the adiabatic motion of the soliton can be deduced from an ordinary (scalar) DNLS; from this, we show that the quantized motion of the soliton is directly related to the quantized displacement of Wannier functions upon pumping, which is known to be set by the Chern number of the band~\cite{asboth2016short,mei2014topological,lohse2016thouless}. This general approach allows us to mathematically demonstrate the topological nature of nonlinear Thouless pumps, by relating the quantized motion of solitons to the Chern number of the underlying Bloch band, see Figs.~\ref{Fig_sketch}(a)-(b). More generally, these developments introduce a theoretical framework by which a broad class of nonlinear topological phenomena can be formulated in terms of topological band indices.

We then broaden the scope by applying this nonlinear topological framework to the realm of quantum gases. By considering an instructive mapping to a Bose-Bose atomic mixture on a lattice~\cite{bloch2008many,chin2010feshbach,1D_mixtures}, which encompasses the aforementioned DNLS as its semiclassical limit, we identify a scenario by which a topological pump  emerges from inter-particle interaction processes:~a soliton of impurity atoms is dragged by the driven majority atoms, hence leading to interaction-induced topological transport; see Fig.~\ref{Fig_sketch}(c). This intriguing phenomenon, which could be implemented in ultracold atomic mixtures in optical lattices~\cite{bloch2008many,chin2010feshbach,1D_mixtures}, is reminiscent of topological polarons~\cite{grusdt2016interferometric,grusdt2019topological,camacho2019dropping,de2020anyonic,pimenov2021topological,PhysRevB.104.035133}, in the sense that impurities inherit the topological properties of their environment through genuine interaction processes.

\subsection{Topological pumps of solitons:~General theory} 

Our theoretical framework concerns a generic class of lattice models governed by the DNLS,
\begin{equation}\label{DNLS_1}
    i \partial_t \phi_{i, \alpha} = \sum_{j, \beta} \, H_{i j}^{\alpha \beta}(t) \, \phi_{j, \beta} - g \, |\phi_{i, \alpha}|^2 \phi_{i, \alpha} \, ,
\end{equation}
where the field $\phi_{i, \alpha}$ is defined at the lattice site $\alpha$ of the $i$th unit cell; $H(t)$ is a time-dependent Hamiltonian matrix, which includes a Thouless pump sequence~\cite{PhysRevB.27.6083,asboth2016short}; and $g\!>\!0$ is the (onsite) nonlinearity strength. Equation~\eqref{DNLS_1} preserves the norm of the field, which we set to \(\sum_{\alpha, i} \, |\phi_{i, \alpha}|^2\!=\!1\), without loss of generality.

An illustrative model, used below to validate the general theory, is provided by the two-band Rice-Mele model~\cite{lohse2016thouless}:~a 1D chain with alternating couplings $J_{1,2}(t)$ and staggered potential $\pm \Delta(t)$ (Appendix). Considering the \emph{nonlinear} Rice-Mele model, Eq.~\eqref{DNLS_1} takes the more explicit form 
\begin{align}
&i \partial_t \phi_{i,1}=J_1(t) \phi_{i,2} + J_2(t)  \phi_{i-1,2}+\Delta(t)\phi_{i,1} - g \, |\phi_{i, 1}|^2 \phi_{i, 1} \notag , \\
&i \partial_t \phi_{i,2}=J_1(t) \phi_{i,1} + J_2(t)  \phi_{i+1,1}-\Delta(t)\phi_{i,2} - g \, |\phi_{i, 2}|^2 \phi_{i, 2} \label{Rice-Mele_NLSE}.
\end{align}
Here, the Thouless pump cycle corresponds to a loop in the parameter space spanned by $(J_2-J_1)$ and $\Delta$, which encircles the origin $(J_1\!=\!J_2, \Delta=0)$; see inset of Fig.~\ref{Fig_two} and Appendix. When $g\!=\!0$, the Bloch bands defined in momentum-time space are associated with a Chern number $C\!=\!\pm1$. This topological invariant is known to determine the quantized displacement for a filled band upon each cycle of the pump~\cite{asboth2016short}.

Our analysis starts by studying the adiabatic evolution associated to the general Eq.~\eqref{DNLS_1}, which is characterized by the period of the pump $T$ (exceeding all other time scales). To simplify notations, we use the multi-index $\vc{i}\!=\!(i,\alpha)$ and write $H_{\vc{i}\vc{j}}\!\equiv\!H_{i j}^{\alpha \beta}(t)$. Introducing the \emph{adiabatic time} \(s\!=\!t/T\), Eq.~\eqref{DNLS_1} takes the form \(i \varepsilon \partial_s \phi_{\vc{i}} = \sum_{\vc{j}} \, H_{\vc{i}\vc{j}}(s) \, \phi_{\vc{j}} - g |\phi_{\vc{i}}|^2 \phi_{\vc{i}}\), where \(\varepsilon=1/T\). The solutions to the adiabatic DNLS can be well approximated by stationary states of the form \(\phi_{\vc{i}} \propto e^{-i \theta_s} \, \varphi_{\vc{i}}\), where \(\theta_s\) is a time-dependent phase factor and \(\varphi_{\vc{i}}\) is an instantaneous solution to the stationary nonlinear Schr\"{o}dinger equation (see Appendix and Refs.~\cite{gang2017adiabatic,carles2011semiclassical})
\begin{equation}\label{DNLS_inst_1}
    \mu_{s} \, \varphi_{\vc{i}} = \sum_{\vc{j}} \, H_{\vc{i}\vc{j}}(s) \, \varphi_{\vc{j}} - g \, |\varphi_{\vc{i}}|^2 \varphi_{\vc{i}} \, ,
\end{equation}
where the instantaneous eigenvalue $\mu_{s}$ explicitly depends on the adiabatic time $s$. 

Equation~\eqref{DNLS_inst_1} admits (bright) solitons as stationary state solutions, which are stable localized structures in the bulk. For sufficiently weak nonlinearity, solitons predominantly occupy the band from which they bifurcate~\cite{szameit2010discrete}, while increasing nonlinearity leads to band mixing. In real space, solitons are immobile without external forcing, and are degenerate modulo a lattice translation set by the translational symmetry of the system. By adiabatically changing the Hamiltonian $H_{\vc{i}\vc{j}}(s)$, a single soliton undergoes smooth deformation, and after one period, it is mapped to the manifold of initial solutions, implying translation by an integer multiple of the unit cell. 
The observations of Ref.~\cite{jurgensen2021quantized} suggest that solitons bifurcating from a single Bloch band undergo a quantized displacement dictated by the Chern number of the band~\cite{PhysRevB.27.6083} over each pump cycle. Demonstrating this intriguing relation between the transport of nonlinear excitations and topological band indices is at the core of the present work.

To elucidate the topological nature of nonlinear pumps, we follow Ref.~\cite{alfimov2002wannier} and represent the solitons of Eq.~\eqref{DNLS_inst_1} in the basis of maximally localized Wannier states,
\begin{equation}\label{soliton_expansion_wannier}
    \varphi_{\vc{i}} = \sum_{n} \, \varphi^{(n)}_{\vc{i}} ,\,  \quad \varphi^{(n)}_{\vc{i}} = \sum_{l} \, a^{(n)}_{l} \, w^{(n)}_{\vc{i}}(l) \, ,
\end{equation}
where the superscript \(n\) denotes the occupied band; the index \(l\) labels the unit cell on which the Wannier state is localized; and all dependence on the adiabatic time $s$ is henceforth implicit. The coefficients \(a^{(n)}_{l}\) obey the analogue of Eq.~\eqref{DNLS_inst_1} in the Wannier representation (Appendix)
\begin{equation}\label{NLSE_wannier}
\begin{split}
    \mu_{s} \, a^{(n)}_{l} & = \sum_{l_1} \, \omega_{l-l_1} \, a^{(n)}_{l_1} \, \\ & - g \sum_{n_1,n_2,n_3} \sum_{l_1,l_2,l_3} W^{(\underline{n})}_{\underline{l}} a^{(n_1)*}_{l_1} \, a^{(n_2)}_{l_2}\, a^{(n_3)}_{l_3} \, ,
\end{split}
\end{equation}
where \(\omega_{l} = 1/N \sum_{\mt{k}} \, \mt{exp}(i \mt{k} \, l) \, \epsilon^{(n)}_{\mt{k}}\) is the Fourier transform of the $n$th Bloch band $\epsilon^{(n)}_{\mt{k}}$ associated with $H_{\vc{i}\vc{j}}(s)$; \(\underline{n}=(n, n_1, n_2, n_3)\), \(\underline{l}=(l, l_1, l_2, l_3)\); and \( W^{(\underline{n})}_{(\underline{l})}\) are the following Wannier overlaps
\begin{equation}\label{W}
    W^{(\underline{n})}_{(\underline{l})} = \sum_{\vc{j}} \, w^{(n)*}_{\vc{j}}(l) \,  w^{(n_1)*}_{\vc{j}}(l_1) \,  w^{(n_2)}_{\vc{j}}(l_2) \,  w^{(n_3)}_{\vc{j}}(l_3) \, .
\end{equation}

The Wannier states of a Bloch band are not unique, as they depend on the gauge choice for the Bloch functions \cite{yu2011equivalent}. Nevertheless, a unique set of maximally localized Wannier functions is provided by the eigenstates of the position operator's projection onto the associated band. Since such Wannier functions are exponentially localized, the contribution to the Wannier overlaps in  Eq.~\eqref{W} from Wannier functions corresponding to different unit cells are negligible. The Wannier overlaps can thus be simplified as \(W^{(\underline{n})}_{\underline{l}}=W^{(\underline{n})} \, \delta_{ll_1} \, \delta_{l_1 l_2} \, \delta_{l_2 l_3}\,\), where $    W^{(\underline{n})} = \sum_{\vc{j}} \, w^{(n)*}_{\vc{j}}(l) \,  w^{(n_1)*}_{\vc{j}}(l) \,  w^{(n_2)}_{\vc{j}}(l) \,  w^{(n_3)}_{\vc{j}}(l)$; we point out that $W^{(\underline{n})}$ does not depend on the index $l$, because of translational invariance.

Moreover, in the regime of weak nonlinearity, the initial state soliton occupies a single band~\cite{jurgensen2021chern,jurgensen2021quantized,alfimov2002wannier}, which allows us to neglect inter-band contributions to Eq.~\eqref{NLSE_wannier}. We note that this simplification holds throughout the  evolution of the pump, during which the soliton adiabatically follows the same band.

Under those realistic assumptions, the Wannier representation of the DNLS reduces to the form
\begin{equation}\label{NLSE_Wannier_simple}
\begin{split}
    \mu_{s} \, a^{(n)}_{l} & = \sum_{l_1} \, \omega_{l-l_1} \, a^{(n)}_{l_1} \, - g W^{(n)} |a^{(n)}_{l}|^2 a^{(n)}_{l} \, ,
\end{split}
\end{equation}
where $W^{(n)} = \sum_{\vc{j}} \, \vert w^{(n)}_{\vc{j}}(l) \vert^4 $. Equation~\eqref{NLSE_Wannier_simple} has the form of a scalar DNLS on a simple lattice with one degree of freedom per unit cell labeled by Wannier indices \(l\), with hopping terms involving nearest and beyond-nearest neighbors. The properties of such scalar DNLS are well established~\cite{kevrekidis2009discrete, kevrekidis2003instabilities, kivshar1993peierls}:~Equation~\eqref{NLSE_Wannier_simple} admits inter-site solitons, with maxima on two adjacent sites, and on-site solitons, with their maximum on a single site. The inter-site solitons are known to be unstable against small perturbations, we thus restrict ourselves to the stable on-site solitons. Crucially, on-site solitons are always peaked around a single site (\(l\)) throughout the pumping cycle, as there is a finite energy (Peierls-Naborro) barrier for delocalization~\cite{kevrekidis2009discrete,kivshar1993peierls}. Interestingly, the Peierls-Naborro barrier plays a role analogous to the ``gap condition" of conventional topological physics, by forbidding transitions to other stable states during the adiabatic time evolution. This observation suggests that solitons are dragged by Wannier states upon pumping, hence exhibiting a quantized displacement in real space established by the Chern number~\cite{asboth2016short,mei2014topological,lohse2016thouless}; see Figs.~\ref{Fig_sketch}(a)-(b).

To firmly prove the topological nature of the nonlinear Thouless pump, we evaluate the solitons center-of-mass displacement after one period $s\!=\!1$ (Appendix)
\begin{align}\label{soliton_COM}
    \Delta \langle \varphi^{(n)}, X \varphi^{(n)} \rangle & = \Delta \langle w^{(n)}(0), X w^{(n)}(0) \rangle \, \\ & + \Delta \sum_{l \neq l'} \, a^{(n)*}_{l'} a^{(n)}_{l}\langle w^{(n)}(l'), X w^{(n)}(l) \rangle,\notag
\end{align}
where \(X\) is the position operator of the lattice; \( \langle f, g \rangle \equiv \sum_{\vc{i}} f^{*}_{\vc{i}} g_{\vc{i}} \) is the inner product of fields on the lattice; and \(\Delta (\cdot) \equiv (\cdot)_{s=1} - (\cdot)_{s=0} \). The first term in Eq.~\eqref{soliton_COM} reflects the displacement of Wannier functions upon one pump cycle, which is known to correspond to the Chern number of the band~\cite{asboth2016short,mei2014topological,lohse2016thouless}; the additional terms displayed on the second line are possible corrections due to the finite overlap of different Wannier states. Importantly, we find that these small interference effects are periodic in time (Appendix), such that these correction terms in Eq.~\eqref{soliton_COM} do not contribute to the solitons center-of-mass displacement over a pump cycle. Altogether, this completes the proof:~the displacement of solitons is indeed quantized according to the Chern number of the band from which they emanate.\\

\subsection{Numerical validation} 

\begin{figure}[t]
\centering
\includegraphics[width=8.6cm]{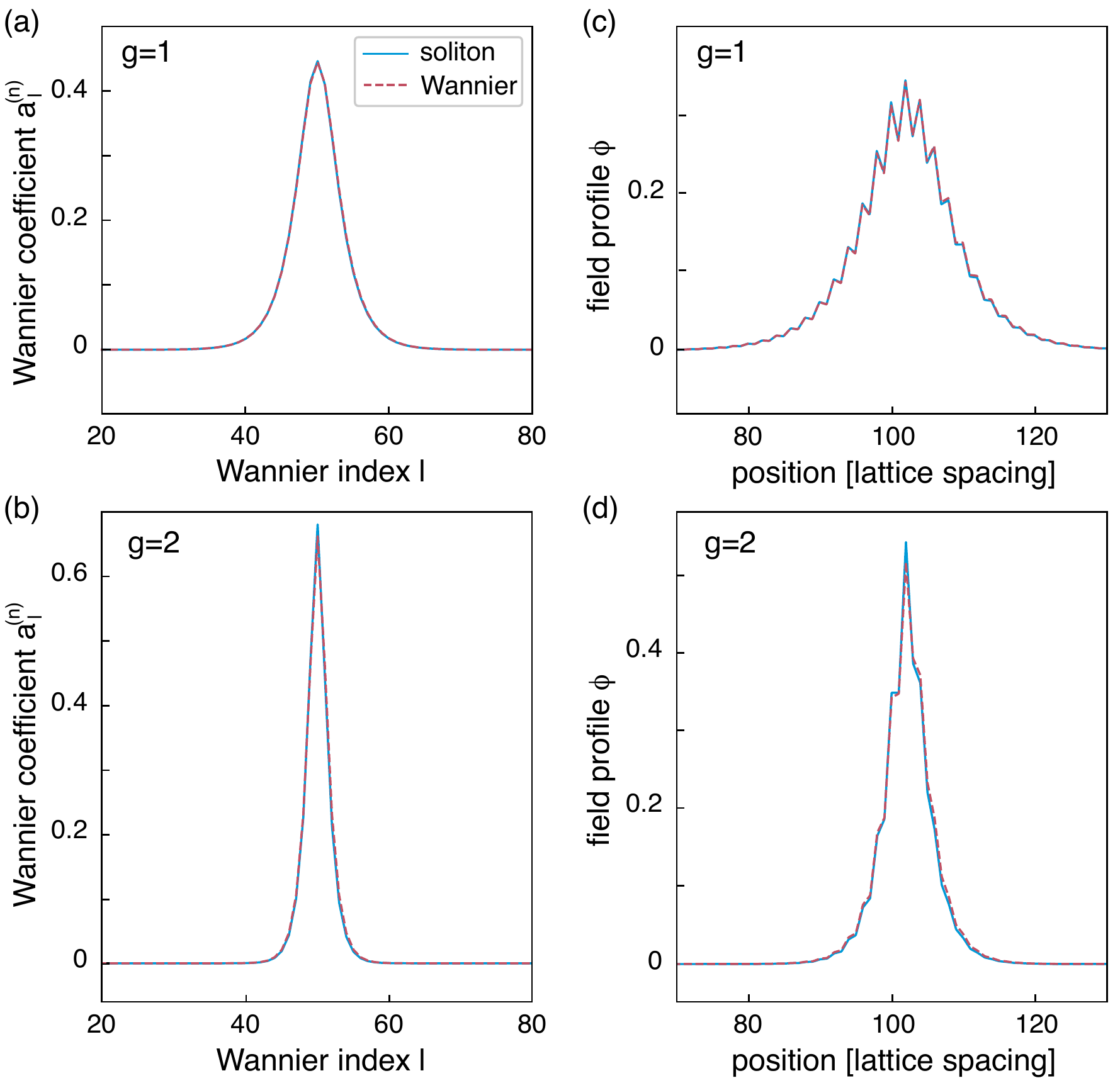}
\vspace{-0.1cm}
\caption{(a)-(b): Wannier representation of a soliton in the lowest band of the nonlinear Rice-Mele model (blue solid line), compared with the soliton obtained from the simplified DNLS Eq.~\eqref{NLSE_Wannier_simple} (dashed red line), for $g\!=\!J_0$ and $g\!=\!2 \, J_0$ and time $s\!=\!0.12$. Here, $J_0$ is a characteristic hopping strength (Appendix). Note how increasing the nonlinearity further localizes the soliton. (c)-(d) Same comparison in real space. }
\label{Fig_one}
\end{figure}

We now demonstrate the validity of our assumptions by solving the nonlinear Rice-Mele model [Eq.~\eqref{Rice-Mele_NLSE}]. In Figs.~\ref{Fig_one}(a)-(b), we compare the on-site soliton solution of the simplified Eq.~\eqref{NLSE_Wannier_simple}, which emerges from the lowest band, with the Wannier representation of the exact soliton obtained by solving the full DNLS in Eq.~\eqref{DNLS_inst_1}. We then perform a similar comparison in real space, by convolving the soliton of Eq.~\eqref{NLSE_Wannier_simple} with the corresponding Wannier states, and by comparing this result to the exact soliton of the original nonlinear Rice-Mele model; see Figs.~\ref{Fig_one}(c)-(d). The perfect agreement validates the description of the soliton in Wannier representation using the ordinary nonlinear Schr\"{o}dinger equation~\eqref{NLSE_Wannier_simple}.

\begin{figure}[h!]
\centering
\includegraphics[width=8.6cm]{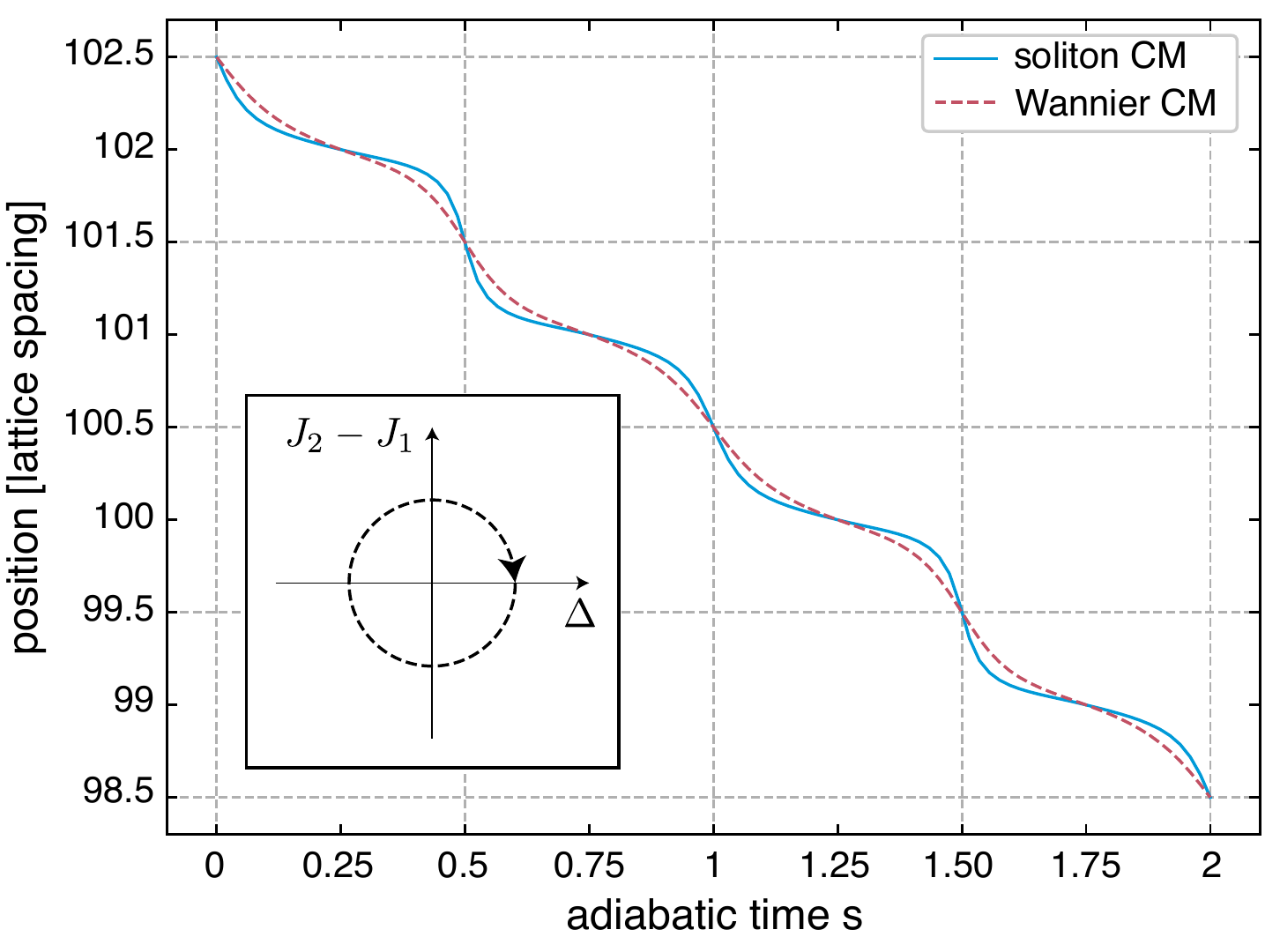}
\vspace{-0.25cm}
\caption{Adiabatic evolution of the soliton's center-of-mass during two full pump cycles (inset), as obtained by solving Eq.~\eqref{DNLS_inst_1} on the Rice-Mele lattice with $g\!=\!J_0$, and selecting a soliton in the lowest band. This is compared to the evolution of the center-of-mass of the Wannier function with largest contribution to the soliton's expansion [Eq.~\eqref{soliton_expansion_wannier}]. For clarity, the Wannier functions are relabeled during the pump cycle such that their center-of-mass follows a contiguous path instead of winding around a unit cell. The quantized displacement is set by the Chern number $C\!=\!-1$ of the occupied band.}
\label{Fig_two}
\end{figure}

We depict the motion of the exact soliton in Fig.~\ref{Fig_two}, as obtained by solving Eq.~\eqref{DNLS_inst_1} over two pump cycles $s\!\in\![0,2]$, and we compare this trajectory with the drift of its underlying Wannier function, i.e.~the Wannier state that contributes the most to the expansion~\eqref{soliton_expansion_wannier}. In order to obtain a contiguous path for the Wannier center, we relabeled the Wannier functions whenever the Wannier centers met discontinuities; this smoothing corresponds to a singular gauge transformation of the corresponding Bloch states, and has no physical implication. Figure~\ref{Fig_two} indicates that the trajectories of the soliton and Wannier center differ at intermediate times ($s\!\ne\!$ integer), which we attribute to the aforementioned interference effects involving different Wannier states (Appendix); however, in agreement with our theoretical predictions, this deviation remains small and time-periodic over the whole pump cycle, and does not introduce any (integer) correction to the quantized center-of-mass displacement.\\

\vspace{-1.715cm}

\subsection{An interaction-induced topological pump \\ for ultracold atomic mixtures} 

The theoretical framework presented in this work is based on the general DNLS in Eq.~\eqref{DNLS_1}, and hence, it applies to a broad range of nonlinear lattice systems. Here, we exploit this universality by introducing a mapping to an imbalanced Bose-Bose atomic mixture, which encompasses the DNLS in Eq.~\eqref{DNLS_1} as its semiclassical limit (within the Thomas-Fermi approximation). As we explain below, this approach reveals an interaction-induced topological pump, where solitons of impurity atoms undergo a quantized drift resulting from genuine inter-particle interaction processes; see Fig.~\ref{Fig_sketch}(c). 

\vspace{-0.11cm}

We start from a microscopic theory for an imbalanced Bose-Bose atomic mixture  on a 1D lattice~\cite{1D_mixtures}, as described by the second-quantized Hamiltonian
\begin{align}
    \hat H &= \sum_{ \langle \vc{i},\vc{j} \rangle} \,    \hat\phi^{\dagger}_{\vc{i}} \, H^{(\phi)}_{\vc{i}\vc{j}} \, \hat\phi_{\vc{j}} 
    + \sum_{\vc{i}} \, \frac{U_{\phi\phi}}{2} \, \hat\phi^{\dagger}_{\vc{i}} \hat\phi^{\dagger}_{\vc{i}} \hat\phi_{\vc{i}} \hat\phi_{\vc{i}} \notag \\
    &+ \sum_{\langle \vc{i}, \vc{j} \rangle} \, \hat\sigma^{\dagger}_{\vc{i}} \, H^{(\sigma)}_{\vc{i}\vc{j}} \, \hat\sigma_{\vc{j}} 
    + \sum_{\vc{i}} \, \frac{U_{\sigma \sigma}}{2} \, \hat\sigma^{\dagger}_{\vc{i}} \hat\sigma^{\dagger}_{\vc{i}} \hat\sigma_{\vc{i}} \,
    \hat\sigma_{\vc{i}} \, \notag \\
    &+ \, \sum_{\vc{i}} \, U_{\phi \sigma} \, \hat\phi^{\dagger}_{\vc{i}}\hat\phi_{\vc{i}} \, \hat\sigma^{\dagger}_{\vc{i}} \hat{\sigma}_{\vc{i}}.\label{eq:imp_bos_Ham}
\end{align}
where $\hat\phi_{\vc{i}}$ and $\hat\sigma_{\vc{i}}$ are bosonic field operators on the lattice; note that we use the same conventions for indices \(\vc{i}=(i,\alpha)\) as before. Specifically, the first line describes single-body processes (i.e.~nearest-neighbor hopping and onsite potentials) and intra-species contact interaction processes for the majority atoms, which are described by the field operator $\hat\phi_{\vc{i}}$; the second line describes single-body processes and intra-species contact interactions for impurity atoms, represented by the field operator $\hat\sigma_{\vc{i}}$; and the third line describes inter-species interaction processes. We assume that the intra-species interaction strengths are both repulsive, $(U_{\sigma \sigma}, U_{\phi \phi} > 0)$, whereas the inter-species interaction strength is attractive (\(U_{\phi \sigma} < 0\)).

In the semi-classical limit, where quantum fluctuations are suppressed for both species, this Bose-Bose mixture setting is well described by two coupled nonlinear Schr\"odinger equations (Appendix and Ref.~\cite{1D_mixtures})
\vspace{-0.2cm}
\begin{equation}    (\omega_0+\mu_{\phi}) \, \phi_{\vc{i}} - \sum_{\vc{j}} \, H^{(\phi)}_{\vc{i}\vc{j}} \, \phi_{\vc{j}} 
    - \Bgpl g_{\phi \phi}|\phi_{\vc{i}}|^2 \, 
    + g_{\phi \sigma}|\sigma_{\vc{i}}|^2 \Bgpr \phi_{\vc{i}} = 0 \, , \notag\\ 
\end{equation}
\vspace{-0.52cm}
\begin{equation}\label{var_eq_12_main}
    (\omega_0+\mu_{\sigma}) \, \sigma_{\vc{i}} - \sum_{\vc{j}} \, H^{(\sigma)}_{\vc{i}\vc{j}} \, \sigma_{\vc{j}} - 
    \Bgpl
    g_{\phi \sigma}|\phi_{\vc{i}}|^2 + g_{\sigma \sigma} \, |\sigma_{\vc{i}}|^2 
    \Bgpr \, \sigma_{\vc{i}} = 0 \, ,
\end{equation}
where $\phi_{\vc{i}}$ and $\sigma_{\vc{i}}$ denote classical fields satisfying the constraints \(\sum_{\vc{i}} \, |\phi_{\vc{i}}|^2 = N_{\phi}/(N_{\phi}+N_{\sigma})\) and \(\sum_{\vc{i}} \, |\sigma_{\vc{i}}|^2 = N_{\sigma}/(N_{\phi}+N_{\sigma})\), with \(N_{\phi}\) and \(N_{\sigma}\)  the particle number of majority and impurity species, respectively; the interaction parameters are defined as $g_{\alpha \beta}\!=\!U_{\alpha \beta} (N_{\phi}+N_{\sigma})$, with $\alpha,\beta=(\phi, \sigma)$; $\mu_{\phi,\sigma}$ denote the chemical potentials; and \(\omega_0\) is the eigenvalue of the nonlinear Eqs.~\eqref{var_eq_12_main}.

Considering the case of heavy impurities, we neglect their kinetic-energy contributions ($H^{(\sigma)}_{\vc{i}\vc{j}}$) to Eq.~\eqref{var_eq_12_main}, the so-called Thomas-Fermi approximation. In this regime, one can relate the impurity mean-field profile to the majority profile as 
\begin{equation}
|\sigma_{\vc{i}}|^2 =- (g_{\phi \sigma}/g_{\sigma \sigma}) \, |\phi_{\vc{i}}|^2,\label{profile_match}
\end{equation}
and Eq.~\eqref{var_eq_12_main} simplifies to the DNLS (Appendix)
\begin{equation}\label{EOM_HS}
    \begin{split}
         (\omega_0+\mu_{\phi}) \,  \phi_{\vc{i}} = \left (\sum_{\vc{j}} \, H^{(\phi)}_{\vc{i}\vc{j}} - u^{\mt{MF}}_{\vc{i}} \right)\phi_{\vc{i}} \, , \quad u^{\mt{MF}}_{\vc{i}} = g |\phi_{\vc{i}}|^2 \, ,
    \end{split}
\end{equation}
where \(g\!=\!-g_{\phi \phi}+g^{2}_{\phi \sigma}/g_{\sigma \sigma}\). Interestingly, Eq.~\eqref{EOM_HS} is formally equivalent to the DNLS in Eq.~\eqref{DNLS_inst_1}:~the majority atoms described by the field $\phi_{\vc{i}}$ can form a soliton and undergo a quantized motion upon driving a Thouless pump sequence in the corresponding lattice Hamiltonian, i.e.~$H^{(\phi)}_{\vc{i}\vc{j}}(s)$. Importantly, according to Eq.~\eqref{profile_match}, the impurity atoms also form a soliton and undergo a quantized motion:~the impurities exhibit topological pumping from genuine interaction processes with the majority atoms. In particular, this interaction-induced topological pumping occurs even when the lattice felt by the impurities $H^{(\sigma)}_{\vc{i}\vc{j}}$ is associated with a trivial (non-topological) band structure.

We first analyze this interaction-induced topological effect by considering the Thomas-Fermi approximation. It appears from Eq.~\eqref{EOM_HS} that $u^{\mt{MF}}$ acts as an effective potential for the majority atoms; a soliton then emerges as the bound state of the impurity field. In the context of highly-imbalanced mixtures with strong impurity-majority coupling, i.e.~in the strong-coupling Bose polaron regime, it is customary to assume a variational ansatz describing the profile of the impurity and majority fields \cite{grusdt2015new}; the majority field is then found as the bound state of the impurity potential \(u^{\mt{MF}}\) using the first relation in Eq.~\eqref{EOM_HS}. Here, the variational problem for obtaining \(u^{\mt{MF}}\) reduces to one for \(\phi\), because of the constraint $u^{\mt{MF}} = g |\phi|^2$. As before, we express \(\phi\) in the Wannier basis, and the variational problem is then solved simultaneously for both \(u^{\mt{MF}}\) and \(\phi\)  using the ansatz \(a_{l}=\eta \, \mt{sech}( \xi \, (l-l_0) ) \) for the Wannier coefficients of \(\phi\). The bound state of the resulting impurity potential $u^{\mt{MF}}\!=\!g |\phi|^2$ then corresponds to the soliton (Appendix). 

\begin{figure}[h!]
\centering
\includegraphics[width=8.7cm]{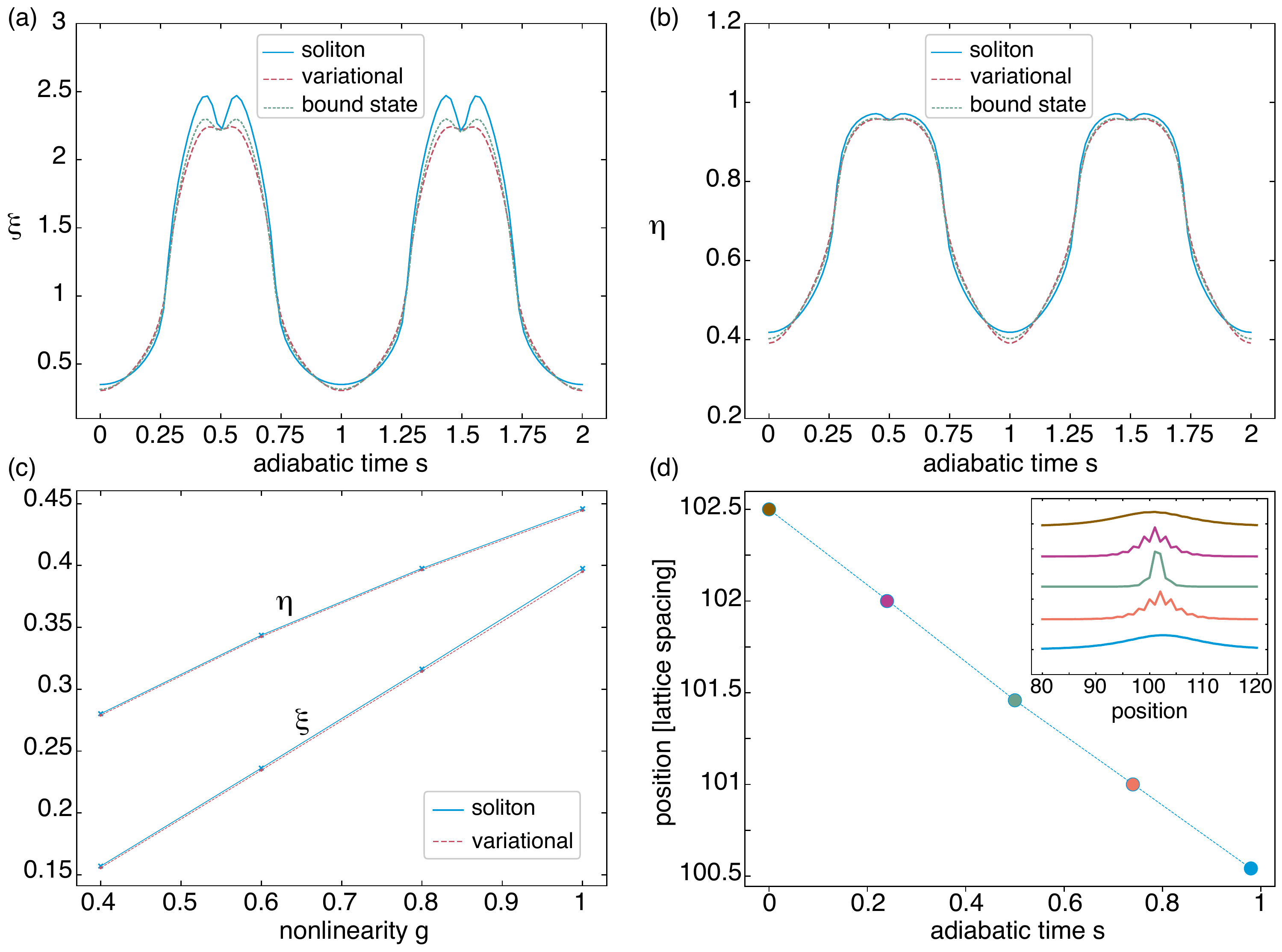}
\vspace{-0.5cm}
\caption{
(a) Evolution of the soliton's width $\xi$ in Wannier space over two pump cycles, as obtained by fitting the numerical solution of Eq.~\eqref{DNLS_inst_1} with a sech function (blue solid line). This is compared to the width of the variational-ansatz solution (dashed red line), and to that of the bound-state solution (green dotted line); here \(g\!=\!J_0\). (b) Same for the amplitude of the soliton $\eta$. (c) Amplitude and width of the exact (solid blue line) and variational-ansatz (dashed red line) solutions as a function of $g$, at time $s\!=\!0.12$. (d) Center-of-mass displacement of the calculated bound state over one pump cycle. The inset shows the corresponding bound state profiles. The quantized motion is dictated by the Chern number $C=-1$; compare with Fig.~\ref{Fig_two}.}
\label{Fig_three}
\end{figure}

Figures~\ref{Fig_three}(a) and (b) show the adiabatic evolution of the amplitude \(\eta\) and width \(\xi\) of the variational solution \(a_{l}\!=\!\eta \, \mt{sech}( \xi \, (l-l_0) ) \) used for the Wannier coefficients of \(\phi\). We compare these results with the amplitude and width extracted from the bound-state solution associated with the impurity potential $u_{\mt{MF}} = g |\phi|^2$, as well as to those extracted from the exact soliton of Eq.~\eqref{DNLS_inst_1} expressed in Wannier representation. We also show the dependence of these parameters on the nonlinearity \(g\) in Fig.~\ref{Fig_three}(c), for both the exact soliton and the variational solution. These results validate our variational approach, as well as the bound-state picture of our soliton.

The minimum-energy solutions obtained from the variational ansatz are realized for integer values of the Wannier index \(l_0\), and thus correspond to stable on-site solitons. Moreover, this Wannier index \(l_0\) remains constant over a pump cycle. Hence, this again suggests that the real-space motion of the soliton should follow the quantized Wannier drift, as established by the Chern number. This is verified in Fig.~\ref{Fig_three}(d), where the center-of-mass displacement of the calculated bound state is shown to be quantized over a pump cycle (compare with Fig.~\ref{Fig_two}).

\begin{figure}[h!]
\centering
\includegraphics[width=8.7cm]{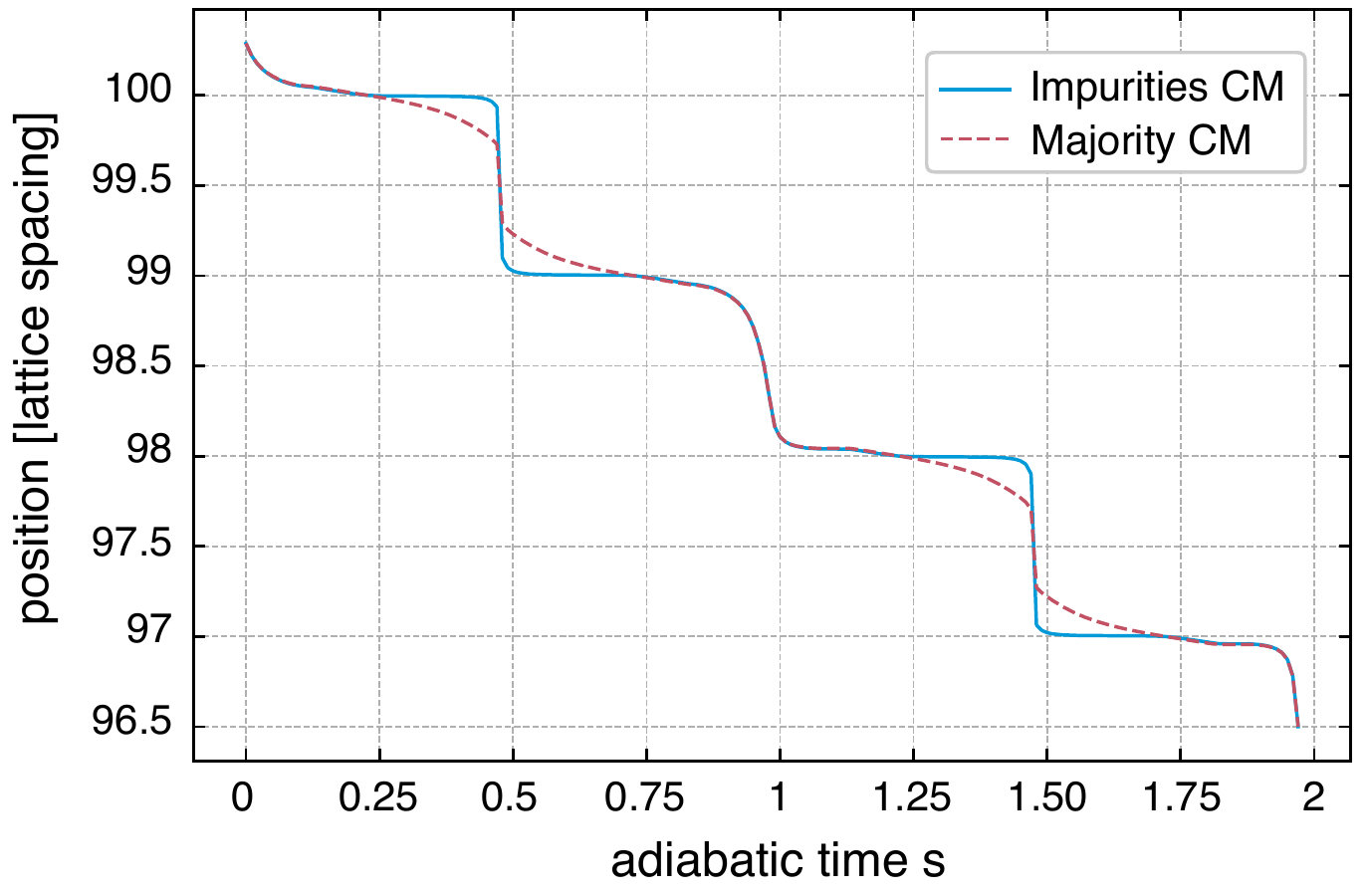}
\caption{Center-of-mass trajectories of both species in a Bose-Bose mixture. Here, the interaction strengths are set as \(g_{\phi \phi} \simeq 0.226 \, J_0, \, g_{\phi \sigma} \simeq -11.32 \, J_0 \) and \(g_{\sigma \sigma} \simeq 2.26 \, J_0\). The majority atoms undergo the pumping sequence as in Fig.~\ref{Fig_two}, while impurities feel a trivial lattice.~Impurity atoms undergo quantized transport through interactions with their environment.}
\label{Fig_pump_mixutre}
\end{figure}
\

In order to demonstrate the validity of our results, in particular, the robustness of the interaction-induced topological pump away from the Thomas-Fermi limit, we solve Eq.~\eqref{var_eq_12_main} numerically for a mass-balanced mixture, thus including the effects of the impurities' kinetic energy. We again use the Rice-Mele model, but consider two different pump sequences for the majority and impurity species:~the majority feels the same (topological) pump sequence as in Fig.~\ref{Fig_two}, while we apply a trivial sequence for the impurity species. We obtain the steady state solution of Eq.~\eqref{var_eq_12_main} over two pump cycles, where the majority particles predominantly occupy the lowest Bloch band. The corresponding trajectories of the CM of both species are depicted in Fig.~\ref{Fig_pump_mixutre}, where the impurity CM is shown to be dragged by the majority particles. While the exact form of the CM trajectories depend on the details of the model and pumping sequence, the CM displacement after one pump cycle is dictated by the Chern number of the topological band occupied by the majority species (\(C\!=\!-1\) in this case). Although the impurity atoms experience a topologically trivial lattice (Appendix), they are shown to undergo topological pumping through genuine interaction effects with their environment. \\

\subsection{Implementation in ultracold atoms}  

The interaction-induced topological pump introduced above could be experimentally implemented in ultracold atomic gases involving two bosonic species. In fact, the  parameters values incorporated in our numerical simulations of Eq.~\eqref{var_eq_12_main}, and displayed in Fig.~\ref{Fig_pump_mixutre}, are compatible with an experimental realization based on bosonic \(^{7}\mt{Li}-^{7}\mt{Li}\) mixtures, with two different hyperfine states of \(^{7}\mt{Li}\) as ``majority" and ``impurity" atoms; we note that the formation of solitons in Lithium gases was previously investigated, both theoretically and experimentally~\cite{ahufinger2004creation,strecker2002formation}. Following Ref.~\cite{hulet2020methods}, the scattering lengths between atoms in state \((F\!=\!1,m_F\!=\!1)\) -- ``impurity" atoms -- and \((F\!=\!1,m_F\!=\!0)\) -- ``majority" atoms -- can be set to \(a_{\phi \phi} \simeq 0.154 \, a_0, \, a_{\phi \sigma} \simeq -7.57 \, a_0, \, a_{\sigma \sigma} \simeq 1.514 \, a_0\), at a magnetic field \(B \simeq 575 \, G\), where \(a_0\) is the Bohr radius (\(a_0=0.0529 \, \mt{nm}\));  we note that these scattering lengths are highly tunable thanks to a broad Feshbach resonance. As further discussed below, this configuration is compatible with the interaction parameters ($g_{\phi \phi}, g_{\sigma \sigma}, g_{\phi \sigma}$) used in our numerics.

The lattice structure and pump sequence can be designed within a time-dependent optical lattice. For instance, following Ref.~\cite{nakajima2016topological},
the atoms can be loaded in a potential landscape comprised of two superimposed optical lattices, with a long-wavelength lattice (\(\lambda_l\!=\!1064 \, \mt{nm}\)) and a shorter lattice (\(\lambda_s\!=\!\lambda_l/2\)), with different amplitudes (\(V_l\!=\!3.0 \, E_R\) and \(V_s\!=\!1.0 \, E_R\), with \(E_R\!=\!h^2/(2m \, \lambda^2_l)\) the recoil energy of the long lattice). Such an optical lattice potential takes the form \(V(x, \phi) = - V_l \, \mt{cos}^2(2\pi x/\lambda_l-\phi) - V_s \, \mt{cos}^2(2\pi x/\lambda_s)\), and it implements the Rice-Mele lattice considered in our numerics:~the Thouless pump sequence is simply realized by sweeping the phase \(\phi\) from \(0\) to \(2 \pi\). The relevant parameters of the Rice-Mele model can be extracted from a tight-binding analysis of the optical lattice potential~\cite{nakajima2016topological}, and the resulting pump sequence is described by the following elliptic path in parameter space: \(((J_1-J_2)/a)^2+(\Delta/b)^2\!=\!1\), with \(a\simeq0.19 \, E_R\) and \(b\simeq0.475 \, E_R\). In our numerics, we choose a closely related pumping sequence with \(a=0.15 \, E_R\) and \(b=0.5 \, E_R\); this choice does not affect our final conclusions, since topological pumping is robust against smooth deformations of the pumping sequence. Finally, to reveal the interaction-induced topological transport for impurities, we propose to implement a trivial pump sequence for that species only [see Fig.~\ref{Fig_pump_mixutre}]; this could be realized by designing a state-dependent optical lattice~\cite{Jaksch_toolbox}, for instance, using the Floquet-engineering scheme of Ref.~\cite{Jotzu_state}.

The particle numbers of the two species can be set to \(N_{\phi}\!\simeq\!1500\) \cite{bradley1997bec} and \(N_{\sigma}/N_{\phi}\!\simeq\!1/30\). With this choice, we obtain the interaction  parameters according to the relation \(g_{\alpha \beta}/E_R\!=\!(N_{\sigma}+N_{\phi})\, \sqrt{8/\pi}\,k_l \, a_{\alpha \beta} (V_s/E_R)^{3/4}\) \cite{bloch2008many}, where \(\alpha, \beta\!=\! (\phi,\sigma)\) and \(k_l\!=\!2\pi/\lambda_l\). Setting the pump parameter \(J_0\!=\!0.5E_R\), the numerical values for the interaction parameters are obtained as \(g_{\phi \phi} \simeq 0.226 \, J_0, \, g_{\phi \sigma} \simeq -11.32 \, \, J_0 \) and \(g_{\sigma \sigma} \simeq 2.26 \, J_0\), which are the values used in our numerical simulations [Fig.~\ref{Fig_pump_mixutre}].

\subsection{Conclusions}

In this work, we outlined a general theoretical framework that connects Bloch band's topology to nonlinear excitations, hence elucidating the topological transport of solitons in the context of nonlinear Thouless pumps. Solitons are stable states of nonlinear lattice systems described by the paradigmatic discrete nonlinear Schr\"{o}dinger equation (DNLS), which is central in describing nonlinear phenomena in a wide range of physical settings, from nonlinear optics and photonics, to ultracold quantum matter, fluid dynamics and plasma physics. In this sense, characterizing the influence of Bloch band's topology on the behavior of the stable states of DNLS is of prime importance. This program is particularly challenging due to the lack of generic theoretical approaches connecting notions of topological physics to nonlinear systems and vice versa. Furthermore, introducing nonlinearities in more sophisticated topological systems, such as higher-dimensional settings, or lattices exhibiting higher-order topology and symmetry-protected features, could lead to exotic phenomena exhibited by the nonlinear modes of the system; see Ref.~\cite{Liberto_orbital} and references therein. By providing a scheme that naturally connects topological indices of band structures to nonlinear excitations, our work opens the door to the exploration of novel nonlinear topological phenomena.

We also illustrated the universality of our approach, by introducing a topological pump for Bose-Bose atomic mixtures, where one species (impurity atoms) experience a quantized drift through genuine interaction processes with the other species (the surrounding majority atoms). Importantly, the impurity atoms inherit the topological properties of their environment through inter-species interactions. We note that such interaction-induced topology has been previously studied in the context of topological polarons, namely, in mixtures with strong population imbalance, where individual topological excitations can bind to mobile impurities~\cite{grusdt2016interferometric,grusdt2019topological,de2020anyonic,PhysRevB.104.035133}. The present scheme extends those concepts to more complex majority-impurity states, such as coupled coherent states within a superfluid phase. We also point out that the proposed scheme can be implemented using available cold-atom technologies, and the quantized transport of impurities can be measured in-situ, using state-selective imaging techniques~\cite{Bloch_Sylvain_review}. Besides, the Chern number characterizing the interaction-induced topological pump could also be directly extracted by interferometry~\cite{grusdt2019topological}.\\

During the preparation of this manuscript, the authors became aware of a related work by M. J\"urgensen and M. C. Rechtsman~\cite{jurgensen2021chern}.\\

\section{Appendix}

{\bf Adiabatic theorem for NLS.} The adiabatic theorem for NLS (both continuous and discrete forms), follows closely the formulation of its linear counterpart~\cite{carles2011semiclassical,gang2017adiabatic}. For a system with a time-dependent Hamiltonian \(H(t)\), which varies on a time scale \(T\) much larger than all the time scales in the problem, the time-dependent NLS takes the following form (see main text)
\begin{equation}\label{NLS_s}
i \varepsilon \, \partial_s \phi = \, H(s) \, \phi - g |\phi|^2 \phi , \, 
\end{equation}
where \(s=t/T\) is the adiabatic time and \(\varepsilon=1/T\) the rate of change. The stationary state solutions of Eq.~\eqref{NLS_s} are of the form 
\begin{equation}\label{ad_eq}
    \phi_s = e^{-i \theta_s} \, \bgpl \varphi_s + \delta \, \varphi_s \bgpr , \,
\end{equation}
where \(\varphi_s\) is the instantaneous solution of the stationary NLS,
\begin{equation}\label{DNLS_inst_1_SM}
    \mu_{s} \, \varphi_s = \, H(s) \, \varphi_s - g \, |\varphi_s|^2 \varphi_s \, ,
\end{equation}
and \(\theta_s = 1/\varepsilon \bgpl \int^{s}_0 ds' \, \mu_{s'} - \gamma_s \bgpr \) is a global phase factor consisting of a dynamical contribution and a Berry phase, and it can be ignored. The correction term \(\delta \, \varphi_s\) accounts for non-adiabatic variations, and for \(\varepsilon \to 0 \), it behaves as \(||\delta \, \varphi|| \sim \varepsilon \,\)
, hence vanishes in the adiabatic limit \(\varepsilon \to 0 \,\). The relevant dynamical information is therefore encoded in the instantaneous solutions of Eq.~\eqref{DNLS_inst_1_SM}.\\

{\bf The Rice-Mele model and pump sequence.} Throughout this work, we illustrate the general concepts and results using the Rice-Mele model, with periodic boundary conditions. This simple two-band model, which is reviewed in some detail below, is known to exhibit a topological (Thouless) pump sequence. 

The Rice-Mele model is a 1D tight-binding model with alternating nearest-neighbor tunneling matrix elements ($J_1$, $J_2$, $J_1$, $J_2$, \dots), and a staggered on-site potential. We denote the two sites within each unit cell by \(\alpha = A, B\) and the unit cells by \(i, \, 0 \leq i \leq N-1\), where \(N\) is the number of unit cells. The hopping matrix element between sites \(A\) and \(B\) within each unit cell (resp. between adjacent unit cells) is written as \(J_1\!=\!-J(1+\delta)\) (resp. \(J_2\!=\!-J(1-\delta)\)) and the magnitude of the staggered potential on site \(A\) (resp.~\(B\)) equals \(\Delta\) (resp.~\(-\Delta\)). The Hamiltonian of the Rice-Mele model thus reads
\begin{align}
H =&  -\sum^{N-1}_{i=0} \, \Bgbl J(1+\delta) \, \dyad{i,A}{i,B}  \\
& + J(1-\delta) \dyad{i, A}{i-1,B}  \Bgbr \notag\\
&+ \frac{\Delta}{2} \,\sum^{N-1}_{i=0} \Bgbl \, \dyad{i,A}{i,A} - \, \dyad{i,B}{i,B} \Bgbr + \text{h.c.} \, \label{RM_Ham}
\end{align}

The simulations shown in the main text were performed on a lattice with \(N=100\) unit cells, and using the following pump sequence 
\begin{equation}
    \begin{split}
        & J(s) = J_0 \bgpl 1 + 1/2 \, \mt{cos}(2\pi s) \bgpr \, ,\\
        & \delta(s) = \delta_0 \, \mt{cos}(2 \pi s)/( 2 + \mt{cos}(2\pi s) ) \, , \\
        & \Delta(s) = \, J_0 \, \mt{sin}(2\pi s), 
    \end{split}\label{sequence}
\end{equation}
with \(J_0\!=\!0.5 \,\) and \(\delta_0\!=\!0.6\), corresponding to a topological pump with Chern number $C\!=\!-1$. The \emph{nonlinear} Rice-Mele model, which is used in our simulations, is obtained by adding an on-site nonlinearity to this lattice model; see Eq.~\eqref{Rice-Mele_NLSE}. 

In order to demonstrate the interaction-induced topological pumping in the Bose-Bose mixture setting, we assume that the two species experience the same Rice-Mele lattice described above, but with different pump sequences:~the majority atoms experience the topological pumping sequence in Eq.~\eqref{sequence}, while the impurity atoms experience a trivial sequence with \(J \delta\!=\!\Delta\!=\!0\). The resulting center-of-mass displacement of both species are depicted in Fig.~\ref{Fig_pump_mixutre} of the main text.\\

{\bf Derivation of the scalar DNLS.} We outline the derivation of the simplified scalar DNLS from the original lattice DNLS,
\begin{equation}\label{NLS_ij}
    \mu \, \phi_{\vc{i}} = \sum_{\vc{j}} \, H_{\vc{i}\vc{j}} \, \phi_{\vc{j}} - g \, |\phi_{\vc{i}}|^2 \, \phi_{\vc{i}} \, .
\end{equation}
The Wannier functions are related to the Bloch waves of the Hamiltonian by the following relations
\begin{equation}\label{w_psi}
\begin{split}
    w^{(n)}_{\vc{j}}(l) & = \frac{1}{\sqrt{N}} \, \sum^{N-1}_{\mt{k}=0} \, e^{i(2\pi/N)\mt{k}(-l)} \, \psi^{(n)}_{\vc{j}}(\mt{k}) \, \\
    & = 
    \frac{1}{\sqrt{N}} \, \sum^{N-1}_{\mt{k}=0} \, e^{i(2\pi/N)\mt{k}(j-l)} \, u^{(n)}_{\vc{j}}(\mt{k}), \,
\end{split}
\end{equation}
where \(\psi^{(n)}_{\vc{j}}(\mt{k})=e^{i(2\pi/N)\mt{k}(j)}\,u^{(n)}_{\vc{j}}(\mt{k})\) is the Bloch wave of band \(n\) with momentum \(\mt{k}\) and \(u^{(n)}_{\vc{j}}(\mt{k})\) is the corresponding Bloch function, which is periodic over the unit cells and does not depend on \(j\). To represent the Hamiltonian part in Wannier basis, we evaluate the matrix elements of the Hamiltonian over the Wannier states
\begin{equation}\label{H_ww}
\begin{split}
    & \langle w^{(n')}(l'), H \, w^{(n)}(l) \rangle  \\ & = \frac{1}{N} \, \sum^{N-1}_{\mt{k}, \mt{k}'=0} \, e^{i(2\pi/N)(\mt{k}'l'-\mt{k}l)} \, \langle \psi^{(n')}(\mt{k}'), \, H \, \psi^{(n)}(\mt{k}) \rangle \\ 
    & =  \delta_{nn'} \, \cdot \, \frac{1}{N} \, \sum^{N-1}_{\mt{k}=0} \, e^{i(2\pi/N)\mt{k}(l'-l)} \, \, \epsilon^{(n)}_{\mt{k}} \, = \delta_{nn'} \, \cdot \, \omega_{l'-l} \, ,
    \end{split}
\end{equation}
where \(\omega_{l} = 1/N \, \sum^{N-1}_{\mt{k}=0} \, e^{i(2\pi/N)\mt{k}(l)} \, \epsilon^{(n)}_{\mt{k}}\) is the Fourier transform of the Bloch band \(\epsilon^{(n)}_{\mt{k}}\); see main text.

Next, we express the nonlinearity in terms of Wannier functions,
\begin{equation}\label{www}
\begin{split}
    & \langle w^{(n)}(l), \, |\phi|^2 \, \phi
    \rangle 
    \\ & = \sum_{n_1, n_2, n_3} \sum_{l_1, l_2, l_3 } \, \\ & 
    \Bggpl \sum_{\vc{i}} \,
    w^{(n)*}_{\vc{i}}(l)w^{(n_1)*}_{\vc{i}}(l_1)w^{(n_2)}_{\vc{i}}(l_2)w^{(n_3)}_{\vc{i}}(l_3) \, \Bggpr
    \, \\ & \times
    a^{(n_1)*}_{l_1} \, a^{(n_2)}_{l_2} \, a^{(n_3)}_{l_3} \, .
\end{split}
\end{equation}
Taking the inner product of Eq.~\eqref{NLS_ij} with \(w^{(n)}_{l}\) and using Eqs.~\eqref{H_ww} and \eqref{www}, we obtain the following DNLS
\begin{equation}\label{NLSE_wannier_SM}
\begin{split}
    \mu_{s} \, a^{(n)}_{l} & = \sum_{l_1} \, \omega_{l-l_1} \, a^{(n)}_{l_1} \, \\ & - g \sum_{n_1,n_2,n_3} \sum_{l_1,l_2,l_3} W^{(\underline{n})}_{\underline{l}} a^{(n_1)*}_{l_1} \, a^{(n_2)}_{l_2}\, a^{(n_3)}_{l_3} \, .
\end{split}
\end{equation}

{\bf Derivation of the soliton center-of-mass displacement.} Here, we prove that the quantized displacement of the solitons center-of-mass is determined by the Chern number of the related Bloch band. For later convenience, we derive the following identity for matrix elements of position operator over the Wannier functions,
\begin{equation}\label{useful_id}
\begin{split}
    & \langle w^{(n)}(l'), X w^{(n)}(l) \rangle \\ & = \langle w^{(n)}({l'-l}), (T^{\dagger}_{l} \, X \, T_{l}) \,  w^{(n)}({0}) \rangle \\ 
    & = \langle w^{(n)}({l'-l}), X \,  w^{(n)}({0}) \rangle + l \, \langle w^{(n)}({l'-l}), \,  w^{(n)}({0}) \rangle \\ 
    & = \langle w^{(n)}({l'-l}), X \,  w^{(n)}({0}) \rangle + l \, \delta_{ll'} 
\end{split}
\end{equation}
where \(T_{l}\) is the translation operator by \(l\) unit cells. In deriving Eq.~\eqref{useful_id} we used the relation \(T^{\dagger}_{l} \, X T_{l} = X + l \,\) together with the orthogonality of Wannier functions. The soliton center-of-mass then reads 
\begin{equation}\label{sol_CM}
\begin{split}
    & \langle
    \varphi^{(n)}, X \varphi^{(n)}
    \rangle_s
    \\ & = 
    \sum_{l,l'} \, 
    a^{(n)*}_{l'} \, a^{(n)}_{l} \,
    \langle 
    w^{(n)}(l'), X w^{(n)}(l)
    \rangle_s \\ 
    & = \sum_{l} \, 
    |a^{(n)}_{l}|^2 \,
    \langle 
    w^{(n)}(l), X w^{(n)}(l) 
    \rangle_s
    \\ & 
    + \sum_{l\neq l'} \, 
    a^{(n)*}_{l'} \, a^{(n)}_{l} \,
    \langle 
    w^{(n)}(l'), X w^{(n)}(l)
    \rangle_s \\
    & = \Bgpl \sum_{l} \, 
    |a^{(n)}_{l}|^2 \Bgpr \,
    \langle 
    w^{(n)}(\mt{0}), X w^{(n)}(\mt{0}) 
    \rangle_s 
    \\ & +
    \Bgpl
    \sum_{l} \, 
    |a^{(n)}_{l}|^2 \,
    l
    \Bgpr
    \, \langle 
    w^{(n)}(\mt{0}), w^{(n)}(\mt{0}) 
    \rangle_s \\
    & + \sum_{\delta l\neq0} \, \Bgpl \sum_{l} \, 
    a^{(n)*}_{l+\delta l} \, a^{(n)}_{l} 
    \Bgpr \,
    \langle 
    w^{(n)}(\delta l), X w^{(n)}(\mt{0})
    \rangle_s \, , 
    \end{split}
\end{equation}
where we used Eq.~\eqref{useful_id} in the last equality. The first term in the last equality of Eq.~\eqref{sol_CM} reduces to \(\langle 
    w^{(n)}(\mt{0}), X w^{(n)}(\mt{0}) 
    \rangle_s\) since we normalized the soliton intensity to unity, \(N_{\phi} = \sum_{l} \, |a^{(n)}_{l}|^2 = 1\). The second term in the last expression is the mean value of the position of the Wannier functions indices, which is constant since the on-site solution is always peaked around a Wannier label and remains symmetric around it. Its contribution to the displacement over a pump cycle thus vanishes. The third term contains products of the form \( \bgpl \sum_{l} \, 
    a^{(n)*}_{l+\delta l} \, a^{(n)}_{l} \bgpr \langle 
    w^{(n)}(\delta l), X w^{(n)}(\mt{0})
    \rangle_s \) and its treatment requires more care. The coefficient \( \bgpl \sum_{l} \, 
    a^{(n)*}_{l+\delta l} \, a^{(n)}_{l} \bgpr \) is time-periodic, since \(a^{(n)}_{l}\) is, by assumption, the solution of the scalar DNLS in Eq.~\eqref{NLSE_Wannier_simple}, in the main text. To investigate the behavior of \(\langle 
    w^{(n)}(\delta l), X w^{(n)}(\mt{0})
    \rangle_s\), we note that after a pump cycle, the Wannier functions are displaced by  the Chern number, \( w^{(n)}(l)|_{s=1} = w^{(n)}(l+\mathcal{C}_n)|_{s=0} \), with \(\mathcal{C}_{n}\) the Chern number of band \(n\). Thus, after a pump cycle, we have 
    \begin{equation}\label{Cross_wan}
    \begin{split}
    \langle 
    & w^{(n)}(\delta l), X w^{(n)}(\mt{0})
    \rangle|_{s=1} \\ & = 
    \langle 
    w^{(n)}(\delta l+\mathcal{C}_n), X w^{(n)}(\mathcal{C}_n)
    \rangle|_{s=0} \\ & = 
    \langle 
    w^{(n)}(\delta l), X w^{(n)}(0)
    \rangle|_{s=0} \, ,
    \end{split}
    \end{equation}
 where we used Eq.~\eqref{useful_id} in the last step. This proves that the quantity \(\langle 
    w^{(n)}(\delta l), X w^{(n)}(0)
    \rangle|_s\), in the last equality of Eq.~\eqref{sol_CM}, is a time-periodic quantity. 
    
    Altogether, the third term in Eq.~\eqref{sol_CM} is also time-periodic, and the soliton's center-of-mass displacement over a pump cycle is given by
    \begin{equation}
        \Delta \langle
    \varphi^{(n)}, X \varphi^{(n)}
    \rangle = \Delta \,
    \langle 
    w^{(n)}(\mt{0}), X w^{(n)}(\mt{0}) 
    \rangle \, .
    \end{equation}
    This result directly relates the soliton's displacement to the displacement of Wannier functions upon one pump cycle, as dictated by the Chern number of the band~\cite{asboth2016short,mei2014topological,lohse2016thouless}. This proves the quantized pumping of the soliton according to the Chern number. \\

{\bf Derivation of the Bose-Bose mixture equations.} In order to derive the equations governing the coherent state profiles of the two species in the mixture, we start from the microscopic Hamiltonian in Eq.~\eqref{eq:imp_bos_Ham}. The coherent-state action of the system  takes the following form (\(\hbar=1\)),
\begin{equation}\label{CS_action}
\begin{split}
    S[\bar{\phi}, \phi; \bar{\sigma}, \sigma] & = \int^{t_f}_{t_i} \, dt \, L[\bar{\phi},\phi;\bar{\sigma},\sigma] \, ,
\end{split}
\end{equation}
with the Lagrangian 
\begin{equation}
\begin{split}
    & L[\bar{\phi},\phi;\bar{\sigma},\sigma] \\ & = 
    \sum_{\vc{i}} \, \bar{\phi}_{\vc{i}} \, 
    \bgbl \, i \partial_t + \mu_{\phi}  
    \bgbr 
    \phi_{\vc{i}}
    - \sum_{\langle \vc{i},\vc{j} \rangle} \, \bar{\phi}_{\vc{i}} \, t_{\phi} H^{(\phi)}_{\vc{i}\vc{j}} \, \phi_{\vc{j}}
    \, - \sum_{\vc{i}} \, \frac{g_{\phi \phi}}{2} \, |\phi_{\vc{i}}|^4 \, 
    \\ &
    +
    \sum_{\vc{i}} 
    \, \bar{\sigma}_{\vc{i}} \,
    \bgbl \, i \partial_t + \mu_{\sigma}  
    \bgbr 
    \sigma_{\vc{i}}
    - \sum_{\langle \vc{i},\vc{j} \rangle} \, \bar{\sigma}_{\vc{i}} \, H^{(\sigma)}_{\vc{i}\vc{j}} \, \sigma_{\vc{j}}
    -  \sum_{\vc{i}} \, \frac{g_{\sigma \sigma}}{2} \, |\sigma_{\vc{i}}|^4 \, \\ & 
    -
    \sum_{\vc{i}} 
    g_{\phi \sigma} \,
    |\sigma_{\vc{i}}|^2 |\phi_{\vc{i}}|^2 .
\end{split}
\end{equation}
To proceed, we seek stationary state solutions for the coherent state fields of the form \(\phi^{(\mt{ss})}_{\vc{i}}(t)=e^{-i\omega_0 t}\,\phi_{\vc{i}}\) and \(\sigma^{(\mt{ss})}_{\vc{i}}(t)=e^{-i\omega_0 t}\,\sigma_{\vc{i}}\), which minimize \( L[\bar{\phi},\phi;\bar{\sigma},\sigma]\). Such solutions are the saddle-point solutions of the quantum mechanical action, giving the mean-field stable states of the system. The Lagrangian then takes the time-independent form
\begin{equation}
\begin{split}
    & L[\bar{\phi},\phi;\bar{\sigma},\sigma] \\ & = 
    \sum_{\vc{i}} \, \bar{\phi}_{\vc{i}} \,
    \bgbl \, \omega_0 + \mu_{\phi}  
    \bgbr 
    \phi_{\vc{i}}
    - \sum_{\langle \vc{i},\vc{j} \rangle} \, \bar{\phi}_{\vc{i}} \, H^{(\phi)}_{\vc{i}\vc{j}} \, \phi_{\vc{j}}
    - \sum_{\vc{i}} \, \frac{g_{\phi \phi}}{2} \, |\phi_{\vc{i}}|^4 \,
    \\ &
    +
    \sum_{\vc{i}} 
    \, \bar{\sigma}_{\vc{i}} \,
    \bgbl \, \omega_0 + \mu_{\sigma}  
    \bgbr 
    \sigma_{\vc{i}}
    - \sum_{\langle \vc{i},\vc{j} \rangle} \, \bar{\sigma}_{\vc{i}} \, H^{(\sigma)}_{\vc{i}\vc{j}} \, \sigma_{\vc{j}}
    -  \sum_{\vc{i}} \, \frac{g_{\sigma \sigma}}{2}\, |\sigma_{\vc{i}} |^4 \, \\ & 
     - \sum_{\vc{i}} g_{\phi \sigma}
    \, |\sigma_{\vc{i}}|^2 \,
    |\phi_{\vc{i}}|^2.
\end{split}
\end{equation}
To minimize the Lagrangian, the corresponding Euler-Lagrange equations are derived from \(\delta L/\delta \, \bar{\phi}_{\vc{i}}=0\, \) and \(\delta L/\delta \, \bar{\sigma}_{\vc{i}}=0\,\), which leads to the two coupled equations in Eq.~\eqref{var_eq_12_main} in the main text. 

In the limiting case of heavy impurities, we neglect their kinetic-energy contributions ($H^{(\sigma)}_{\vc{i}\vc{j}}$) to Eq.~\eqref{var_eq_12_main}, the so-called Thomas-Fermi approximation. In this case, the second equation in Eq.~\eqref{var_eq_12_main} reduces to \( (\omega_0+\mu_{\sigma}) = g_{\phi \sigma}|\phi_{\vc{i}}|^2 + g_{\sigma \sigma} \, |\sigma_{\vc{i}}|^2 \). For the bright soliton solutions of Eq.~\eqref{var_eq_12_main}, \(\phi_{\vc{i}}\) and \( \sigma_{\vc{i}} \) decay exponentially away from the soliton center, thus, to zeroth order in the impurities hopping strength, \(\omega_0+\mu_{\sigma}=0\). Eq.~\eqref{var_eq_12_main} then reduce to
\begin{equation}\label{var_eq_21}
    (\omega_0+\mu_{\phi}) \, \phi_{\vc{i}} = \sum_{\vc{j}} \, H^{(\phi)}_{\vc{i}\vc{j}} \, \phi_{\vc{j}} 
    + \Bgpl g_{\phi \phi}|\phi_{\vc{i}}|^2 \, 
    + g_{\phi \sigma}|\sigma_{\vc{i}}|^2 \Bgpr \phi_{\vc{i}} \, , \\ 
\end{equation}
\begin{equation}\label{var_eq_22}
    |\sigma_{\vc{i}}|^2 = -g_{\phi \sigma}/g_{\sigma\sigma} \, |\phi_{\vc{i}}|^2  \, .
\end{equation}
Inserting Eq.~\eqref{var_eq_22} into Eq.~\eqref{var_eq_21}, we obtain an effective DNLS for \(\phi_{\vc{i}}\),
\begin{equation}\label{var_eq_23}
    (\omega_0+\mu_{\phi}) \, \phi_{\vc{i}} = \sum_{\vc{j}} \, H^{(\phi)}_{\vc{i}\vc{j}} \, \phi_{\vc{j}} 
    + \bgpl g_{\phi \phi} \, 
    - g^2_{\phi \sigma}/g_{\sigma \sigma} \bgpr |\phi_{\vc{i}}|^2 \phi_{\vc{i}} \, , \\ 
\end{equation}
with the effective nonlinearity strength \(g = -g_{\phi \phi} \, 
+ g^2_{\phi \sigma}/g_{\sigma \sigma} \,\), which for \(g_{\phi \phi} g_{\sigma \sigma} < g^2_{\phi \sigma}\) corresponds to a defocusing nonlinearity. \\

{\bf Variational ansatz for the state of Bose-Bose mixture in the Thomas-Fermi limit.} The variational treatment of Eqs.~\eqref{profile_match} and \eqref{EOM_HS} accounts to minimizing the following energy functional for the field \(\phi\)
\begin{equation}\label{funcnl_GPE}
\begin{split}
    H[\bar{\phi}, \phi] & = \sum_{\vc{i},\vc{j}} \, \bar{\phi}_{\vc{i}} \, H^{(\phi)}_{\vc{i}\vc{j}} \, \phi_{\vc{j}} - \frac{g}{2} \sum_{\vc{i}} \,  \vert\phi_{\vc{i}}\vert^2 \\ & - \mu_{\phi} \, \Bgpl \sum_{\vc{i}} \, \vert\phi_{\vc{i}}\vert^2 - N_{\phi}\Bgpr \, .  
\end{split}
\end{equation}
From the knowledge obtained from the soliton solutions of the DNLS in the main text, we assume that \(\phi_{\vc{i}}\) belongs to a single band  and expand it in terms of the Wannier functions of the corresponding band, \(\phi_{\vc{i}} \!=\! \sum_{l} \, a^{(n)}_{l} \, w^{(n)}(l)\). We then use a sech variational ansatz for the coefficient amplitudes, \(a^{(n)}_{l} \!=\! \eta/ \, \mt{sech}\bgpl \xi(l-l_0) \bgpr \). The variational energy functional takes the following form 
\begin{equation}\label{H_var}
\begin{split}
H/\eta^2 & = \omega_0 \, N_{\phi}/\eta^2 + \sum^{\infty}_{n=1} \, \frac{4n}{\mt{sinh}(\xi n)} \omega_n \\ & - \frac{2 \,g}{3} \eta^2 \Bggbl \frac{1}{\xi} + \sum^{\infty}_{m=1} \frac{2 \pi^2}{ \xi^2} \Bgpl 1 + \frac{\pi^2 m^2}{\xi^2} \Bgpr \frac{m \, \mt{cos}(2\pi m l_0)}{\mt{sinh(\frac{\pi^2 m}{\xi})}} \Bggbr \, ,
\end{split}
\end{equation}
subject to the constraint \(N_{\phi}=\mt{const.}\), where
\begin{equation}\label{N_var}
    N_{\phi}/\eta^2 = \frac{2}{\xi} + \sum^{\infty}_{m=1} \, \frac{4\pi^2}{\xi^2} \, \frac{m \, \mt{cos}(2 \pi m l_0)}{\mt{sinh}(\frac{\pi^2 m}{\xi})} \, .
\end{equation}
For the simulations presented in the main text [Fig.~\ref{Fig_three}], we assume that \(N_{\phi}\!=\!1 \,\); see Refs. ~\cite{kevrekidis2009discrete,kevrekidis2003instabilities} for more details on variational ans\"{a}tze for DNLS. From the solution of Eqs.~\eqref{H_var} and \eqref{N_var} we then obtain the boson field, \(\phi_{\vc{i}}\), which is then used to obtain the effective attractive potential \(u_{\vc{i}}^{\text{MF}}=g|\phi_{\vc{i}}|^2\,\); see Eq.~\eqref{EOM_HS}. 


\vspace{0.5cm}

\paragraph{Acknowledgement.}
We are glad to thank B.~Oblak, A. Bedroya, N. Englebert, S.-P. Gorza, F. Leo and S.~Mukherjee for useful discussions. We also acknowledge M.~J\"urgensen and M. C. Rechtsman for fruitful discussions, for sharing their results in the course of our numerical studies and for enlightening us on the negligible role of band mixing. Work in Brussels is supported by the FRS-FNRS (Belgium) and the ERC Starting Grant TopoCold. The authors acknowledge funding by the Deutsche Forschungsgemeinschaft (DFG, German Research Foundation) under Germany's Excellence Strategy -- EXC-2111 -- 390814868 and via Research Unit FOR 2414 under project number 277974659.

\clearpage

\end{document}